\documentclass[lettersize,journal]{IEEEtran}
\usepackage{amsmath,amsfonts}
\usepackage{algorithmic}
\usepackage{algorithm}
\usepackage{array}
\usepackage[caption=false,font=normalsize,labelfont=sf,textfont=sf]{subfig}
\usepackage{textcomp}
\usepackage{stfloats}
\usepackage{url}
\usepackage{verbatim}
\usepackage{graphicx}
\usepackage{cite}
\usepackage{pifont}
\usepackage{multirow}

\usepackage{amssymb} 
\usepackage{float}
\usepackage{adjustbox}
\usepackage{hyperref}
\usepackage{cleveref}
\usepackage{color}

\begin{document}

\title{FNSE-SBGAN: Far-field Speech Enhancement with Schrödinger Bridge and Generative Adversarial Networks}

\author{Tong Lei,~\IEEEmembership{Student Member,~IEEE,} Qinwen Hu, Ziyao Lin, Andong Li,~\IEEEmembership{Member,~IEEE,} Rilin Chen, Meng Yu,~\IEEEmembership{Member,~IEEE,} Dong Yu,~\IEEEmembership{Fellow,~IEEE,} and Jing Lu,~\IEEEmembership{Member,~IEEE}
\thanks{Manuscript received ----, 2025; revised ----, 2025. This work was supported by the National Natural Science Foundation of China (Grant No. 12274221).}
\thanks{Tong Lei, Qinwen Hu, Ziyao Lin and Jing Lu are with the Key Laboratory of Modern Acoustics, Nanjing University, Nanjing, 210093, Jiangsu, China. Jing Lu is also with NJU-Horizon Intelligent Audio Lab, Horizon Robotics, Beijing 100094, China. This work was done while Tong Lei was a research intern at Tencent AI Lab, Beijing, China. (\textit{Corresponding author: Jing Lu})}
\thanks{Andong Li is with the Key Laboratory of Noise and Vibration Research, Institute of Acoustics, Chinese Academy of Sciences, Beijing, 100190, China. }
\thanks{Tong Lei and Rilin Chen are with Tencent AI Lab, Beijing, China.}
\thanks{Dong Yu and Meng Yu are with Tencent AI Lab, Bellevue, WA, USA.}
\thanks{(\textit{Email: tonglei@smail.nju.edu.cn; qinwen.hu@smail.nju.edu.cn; liandong@mail.ioa.ac.cn; ziyaolin@smail.nju.edu.cn; rilinchen@-tencent.com; raymondmyu@global.tencent.com; dyu@global.tencent.com; lujing@nju.edu.cn})}
}

\markboth{Journal of \LaTeX\ Class Files,~Vol.~14, No.~8, August~2021}%
{Shell \MakeLowercase{\textit{et al.}}: A Sample Article Using IEEEtran.cls for IEEE Journals}


\maketitle

\begin{abstract}
The prevailing method for neural speech enhancement predominantly utilizes fully-supervised deep learning with simulated pairs of far-field noisy-reverberant speech and clean speech. Nonetheless, these models frequently demonstrate restricted generalizability to mixtures recorded in real-world conditions. To address this issue, this study investigates training enhancement models directly on real mixtures. Specifically, we revisit the single-channel far-field to near-field speech enhancement (FNSE) task, focusing on real-world data characterized by low signal-to-noise ratio (SNR), high reverberation, and mid-to-high frequency attenuation. We propose FNSE-SBGAN, a framework that integrates a Schrödinger Bridge (SB)-based diffusion model with generative adversarial networks (GANs). Our approach achieves state-of-the-art performance across various metrics and subjective evaluations, significantly reducing the character error rate (CER) by up to 14.58\% compared to far-field signals. Experimental results demonstrate that FNSE-SBGAN preserves superior subjective quality and establishes a new benchmark for real-world far-field speech enhancement. Additionally, we introduce an evaluation framework leveraging matrix rank analysis in the time-frequency domain, providing systematic insights into model performance and revealing the strengths and weaknesses of different generative methods.
\end{abstract}


\section{Introduction}
\IEEEPARstart{S}{peech} enhancement (SE) aims to extract clean speech while suppressing background noise, with the objectives of enhancing speech intelligibility and perceptual quality. It plays a critical role in speech frontend processing and various applications, such as voice recognition~\cite{wang2020complex,gnanamanickam2021hybrid}, telecommunication systems~\cite{robustSE}, hearing aids~\cite{park2020speech,hearingaids}, and assistive technologies for individuals with hearing impairments~\cite{tools-hearimpair}. Over the last decade, the data-driven deep learning (DL)-based models have been extensively employed for SE and have demonstrated superior performance compared to conventional rule-based signal processing methods~\cite{wang2018supervised}.

Far-field speech enhancement is confronted with distinctive challenges stemming from the aggravated complexity of acoustic environments, characterized by elevated reverberation levels, diminished signal-to-noise ratios (SNR), and increased variability in the directivity patterns of microphones and sound sources. These factors collectively contribute to a more demanding task of achieving robust performance in real-world scenarios when contrasted with near-field conditions. As delineated in our prior research~\cite{LEI:FNSESAT}, the training paradigm for deep learning-based models typically involves the convolution of near-field acoustic signals with two categories of room impulse responses (RIRs): RIRs obtained through physical measurements or RIRs synthesized via the image source method (ISM)~\cite{image-small-room} in conjunction with the diffusion field method~\cite{habets2008generating}. Subsequently, these convolved signals are superimposed with diverse noise types at predetermined SNR levels to construct simulated datasets. Nonetheless, substantial discrepancies persist between the acoustic attributes of simulated data and real-world recordings, particularly in the following aspects:

\subsubsection{Directivity Characteristics and Absorption Properties} 
The ISM framework~\cite{image-small-room} presupposes that both microphones and sound sources exhibit omni-directional radiation patterns, whereas practical scenarios reveal that microphones, especially when affixed to specific devices, demonstrate discernible directivity, and human speakers predominantly exhibit frontal (on-axis) directivity~\cite{brandner2018dirpat—database}. Moreover, ISM conventionally employs frequency-independent wall absorption coefficients, which inadequately capture the behavior of actual wall materials. The incorporation of frequency-dependent microphone directivity, sound source directivity, and wall absorption coefficients into ISM, as demonstrated in~\cite{Realistic-Sources}, markedly enhances the verisimilitude of simulated RIRs.

\subsubsection{Spatial Configuration and Furnishing Elements}
ISM predominantly simulates acoustically empty, rectangular rooms, in contrast to real-world environments that often feature irregular geometries and the presence of furniture. To emulate such irregular spaces and the scattering effects induced by furnishings, more sophisticated geometric acoustic techniques, such as ray-tracing methods~\cite{GWA-dataset}, can be employed, albeit at increasing computational expense. The task of configuring room geometry, spatial layout, and the material properties of both the room and its furnishings to align with real-world scene semantics presents an additional layer of complexity. This challenge is effectively addressed in~\cite{GWA-dataset} through the utilization of scene computer-aided design (CAD) models for room configuration and the application of natural language processing techniques for material specification, thereby substantially augmenting the authenticity of simulated RIRs. In addition, Li \emph{et al.}~\cite{likai2024sonicsim} introduce the SonicSim, a synthetic toolkit designed to generate highly customizable data for moving sound sources, which can be used to construct a benchmark dataset for moving sound sources and to evaluate the performance of speech separation and enhancement models.

Despite the significant advancements in simulation tools, they remain challenging to capture the full complexities of real-world data. This limitation has spurred growing research interest in leveraging real-world data directly, leading to notable progress in recent years. A persistent challenge in this area lies in obtaining paired clean close-talk targets for training on far-field signals. 
%
%
To address this, RealMan~\cite{yang2024realman} captures source speech and microphone array signals in dynamic acoustic environments using synchronized recording equipment, creating the first real-world far-field dataset that supports direct-path speech estimation. Specifically, the direct-path target clean speech is derived by filtering the source speech with estimated direct-path impulse responses.
In contrast, ctPuLSE~\cite{wang2024ctpulse} proposes a pseudo-label based approach for far-field speech enhancement, where the estimated close-talk speech serves as a pseudo-supervised target for training supervised enhancement models directly based on real far-field mixtures, thereby potentially leading to better generalization capability to real-world data.

In our previous work FNSE-SAT~\cite{LEI:FNSESAT}, our approach involves using synchronized recording equipment to capture far-field and near-field speech in various indoor and outdoor scenarios for DL model training. The near-field speech is pre-processed to remove slight noise and reverberations. In this context, SE aims at recovering near-field speech from far-field speech, addressing the challenges posed by the degradation of speech quality in far-field conditions. Typically, this recovery task requires using generative models, which leverage the prior distribution of clean near-field speech to avoid issues such as ``regression to the mean" and ``over-suppression" problems~\cite{lemercier2023storm}.  In FNSE-SAT~\cite{LEI:FNSESAT}, we explore supervised adversarial training with a focus on generative adversarial networks (GANs) to enhance real-world single-channel far-field speech signals. While this approach significantly enhances speech quality and intelligibility of far-field signals, it still faces challenges in achieving excellent naturalness in the reconstructed signals. Therefore, it is necessary and also significant to further incorporate the advancements in generative models.

Recent breakthroughs in generative models have revolutionized speech-related domains including text-to-speech synthesis (TTS)~\cite{wang:tacotron,ren:fastspeech2,ren:fastspeech,tan:naturespeech}, text-to-audio (TTA)~\cite{huang:MAA-TTA,majumder:tango2}, voice conversion~\cite{qian:autovc,choi:neural}, audio editing~\cite{wang:audit}, and singing voice synthesis (SVS)~\cite{liu:diffsinger,hwang:hiddensinger}. In addition to GANs~\cite{saxena2021generative,wali2022generative}, diffusion-based generative models have also made significant progress. Score-Based Generative Models (SGMs) excel at traversing complex data manifolds through stochastic differential equations~\cite{song:scorebased}, making them robust to low-SNR conditions~\cite{ho2020denoising}. Flow-Based Architectures (RealNVP~\cite{dinh2017density}, Glow~\cite{kingma2018glow}, Neural ODEs~\cite{chen2018neural} and etc.) offer invertible transformations that preserve phonetic content during enhancement~\cite{prenger:waveglow}. Flow Matching Techniques (Rectified Flow Matching, Optimal-Transport Flow Matching and etc.) provide efficient training through direct vector field alignment, particularly effective for real-world data distributions~\cite{lipman2023flow}. Emerging extensions like Schrödinger Bridge (SB) models further bridge these approaches by enabling direct estimation of optimal transport paths between degraded and clean speech distributions~\cite{de2021diffusion}. This capability is critical for addressing the tripartite challenges of real-world far-field speech: spectral attenuation~\cite{blackstock2000fundamentals}, reverberation artifacts~\cite{naylor2010speech}, and nonlinear phase distortions~\cite{image-small-room}.

In this paper, considering the distinctive characteristics of real-recorded far-field data, which include: 1) low SNR, 2) high reverberation levels, and 3) significant attenuation of mid-to-high frequency components~\cite{LEI:FNSESAT}, we propose to employ the diffusion model, a state-of-the-art (SOTA) generative approach, for the restoration of attenuated and degraded speech information. Specifically, we adopt the Schrödinger Bridge (SB) framework due to its exceptional capability in directly estimating the target speech from intermediate states. Building on this framework, we propose a Far-field to Near-field Speech Enhancement system using Schrödinger Bridge and Generative Adversarial Networks (FNSE-SBGAN). This approach enables the integration of auxiliary loss functions to effectively constrain the generated speech content, minimizing the risk of semantic distortion. The proposed framework demonstrates superior performance, not only maintaining excellent subjective speech quality but also achieving a lower Character Error Rate (CER) compared to both conventional two-stage diffusion methods~\cite{lemercier2023storm} and the FNSE-SAT~\cite{LEI:FNSESAT} approach. 

Furthermore, we revisit the single-channel real-world FNSE task. By conducting an in-depth analysis of the real-world dataset~\cite{LEI:FNSESAT}, we uncover the matrix rank trends of far- and near-field signals in the time-frequency (T-F) domain. Additionally, we measure the rank differences between the outputs of various models and the cleaned near-field signals. Based on this analysis, we establish a correlation between the overall performance of the models and the variance of the rank differences.

The contributions of this paper can be summarized as follows:
\begin{itemize}
\item To the best of our knowledge, FNSE-SBGAN is the first diffusion model-based framework specifically developed to enhance real-world far-field speech signals.
\item By combining a Schrödinger Bridge-based diffusion model with a generative adversarial network, the proposed FNSE-SBGAN method achieves state-of-the-art performance across multiple reference-based and reference-free metrics, as well as in character error rate.
\item We introduce an approach that uses matrix rank to evaluate the effectiveness of models in the task of far-field to near-field speech enhancement.
\item Within the diffusion framework, we conduct experiments and comparisons with various one-stage sampling methods and two-stage speech enhancement methods.
\end{itemize}

The rest of this paper is organized as follows: Section~\ref{sec:promFormulation} reviews the problem formulation, including signal models and rank calculation of the signals. Section~\ref{sec:bkgd} provides the related background, including score-based generative models and flow matching techniques in generative models. Section~\ref{sec:FNSE-SBGAN} details our proposed FNSE-SBGAN framework, explaining its architecture and the integration of Schrödinger Bridge and GAN, and provides a detailed description of the training loss functions. Section~\ref{sec:result} presents the experimental results and analysis, demonstrating the effectiveness of our approach through various metrics and comparisons with existing methods. Finally, Section~\ref{sec:conclusions} concludes the key findings.

\section{\label{sec:promFormulation} Problem Formulation}
\subsection{\label{subsec:Signal-Models} Signal Models}
For real-recorded far-field data, modeling needs to consider not only noise and reverberation but also the distortion caused by the distance from the microphone to the sound source and the nonlinear attenuation due to air absorption:
\begin{equation}
    M_{l,f}=k_fX_{l,f}+H_{l,f}+N_{l,f},
    \label{eq:signal}
\end{equation}
where $\{M_{l,f},X_{l,f},H_{l,f},N_{l,f}\}\in \mathbb{C}$  denote the mixture, source with early reflections, late reverberation, and noise signals, respectively. For simplicity, we refer to the sources with early reflections as source signals hereafter. $k_f$ is the mapping parameter from the near-field measured data to the far-field recorded data. The subscripts $l \in \{1,...,L\}$ and $f\in \{1,...,F\}$ represent the time and frequency indices, respectively. Our goal is to obtain the source signals $X_{l,f}$ from the mixture signals $M_{l,f}$. In fact, real-recorded near-field also contain late reverberation and a small amount of noise signal. To address this, we mitigate the background noise and reverberation in the near-field signals with the pretrained DPCRN model~\cite{le2022inference}, which is labeled as ``Cleaned near-field". After this processing, the ``Cleaned near-field" can be considered as the source signal with early reflections.
\subsection{\label{subsec:rank-analysis} Rank Calculation}
Considering the spectrogram matrices constructed from the magnitude STFT coefficients in Eq.~(\ref{eq:signal}), let $\mathcal{R}(\cdot):\mathbb{R}^{L\times F}\to\mathbb{Z}$ denote the matrix rank operator~{\cite{li2025neural}}. Applying subadditivity of matrix rank in spectral magnitude domain yields
\begin{equation}
    \mathcal{R}(|\mathbf{M}|)\approx\mathcal{R}(|\mathbf{X}\mathbf{k}|+|\mathbf{H}|+|\mathbf{N}|)\leq\mathcal{R}(|\mathbf{X}\mathbf{k}|)+\mathcal{R}(\mathbf{H})+\mathcal{R}(|\mathbf{N}|),
    \label{eq:denoise-rank}
\end{equation}
where $|\mathbf{M}|$, $|\mathbf{X}\mathbf{k}|$, $|\mathbf{H}|$ and $|\mathbf{N}|$ represent magnitude spectrograms. In Eq.~(\ref{eq:denoise-rank}), the phase component is omitted because the rank is mainly associated with the magnitude of matrix elements, which are indicative of signal energy. It provides an upper bound on the rank of the mixture spectrum \(\mathbf{M}\). This implies that after adding noise \(\mathbf{N}\) and late reverberation $\mathbf{H}$, the upper bound of the matrix rank tends to increase.

\section{\label{sec:bkgd} Preliminaries}
We begin with a brief overview of the commonly used diffusion models, specifically score-based generative models (SGMs), including the forward and reverse stochastic differential equations (SDE) and the score matching objective of the score network. 

\subsection{\label{subsec:3:1} Score-Based Generative Models}

We reshape the source signals $\mathbf{X}$ into a vector $\boldsymbol{x}_0=\mathbf{f}_{reshape}(\mathbf{X})$ using a reshaping transformation $\mathbf{f}_{reshape}(\cdot)$,
where $\boldsymbol{x}_0 \in \mathbb{R}^{d}$ and $d=2\times L \times F$. 
Here, $\boldsymbol{x}_0$ and $\boldsymbol{x}_T$ represent the source signal and far-field signal within a $d$-dimensional data space. We assume the maximum time \(T=1\) for convenience. 
In the following formulation, we model the distribution of a variable $\boldsymbol{x}$ that resides in the same data space as $\boldsymbol{x}_0$.
Given a data distribution \(p_{\mathrm{data}}(\boldsymbol{x})\), SGMs~\cite{song:scorebased} are built on a continuous-time diffusion process defined by a forward SDE:
    \begin{equation}
         \mathrm{d}\boldsymbol{x}_t=\boldsymbol{f}(\boldsymbol{x}_t,t)\mathrm{d}t+g(t)\mathrm{d}\boldsymbol{w}_t,\quad\boldsymbol{x}_0\sim p_0=p_\mathrm{data},
         \label{eq:sde}
    \end{equation}
where \(t \in [0,1]\) is a continuous time variable,  \(\boldsymbol{x}_t \in \mathbb{R}^d\) is the state of the process,  \(\boldsymbol{f}\) is a vector-valued drift coefficient, \(g\) is a scalar-valued diffusion coefficient, and \(\boldsymbol{w}_t \in \mathbb{R}^d\) is a standard Wiener process. To ensure that the boundary distribution is a Gaussian prior distribution \(p_{\mathrm{prior}} = \mathcal{N}(\mathbf{0}, \sigma_1^2 \boldsymbol{I})\), we construct the drift coefficient \(\boldsymbol{f}\) and the diffusion coefficient \(g\) accordingly. This construction guarantees that the forward SDE has a corresponding reverse SDE:
    \begin{align}
        \mathrm{d}\boldsymbol{x}_t=[\boldsymbol{f}(\boldsymbol{x}_t,t)-g^2(t)\nabla\log p_t(\boldsymbol{x}_t)]\mathrm{d}t+g(t)\mathrm{d}\bar{\boldsymbol{w}}_t,\nonumber \\
        \boldsymbol{x}_1\sim p_1\approx p_\mathrm{prior},
        \label{eq:revsde}
    \end{align}
where \(\bar{\boldsymbol{w}}_t\) is the reverse-time Wiener process, and $\mathrm{d}t$ denotes an infinitesimal negative time step. \(\nabla\log p_t(\boldsymbol{x}_t)\) is the \textit{score function} of the marginal distribution \(p_t\). The solution trajectories of this reverse SDE share the same marginal densities as those of the forward SDE, except that they evolve in the opposite time direction~\cite{song:scorebased,anderson1982reverse}. Intuitively, solutions to the reverse-time SDE are diffusion processes that gradually convert noise to data. 
    

To enable data sample generation during inference, we can replace the score function with a score network \(s_\theta(\boldsymbol{x}_t,t)\) and solve it reversely from \(p_\mathrm{prior}\) at \(t=1\). The loss function for SGM is in the form of the mean-square error (MSE)~\cite{bishop2006pattern}, and can be expressed as follows~\cite{song:scorebased,vincent2011connection}: 
\begin{equation}
\mathcal{L}_{\text{SGM}} = \mathbb{E}_{t, \boldsymbol{x}_0, \boldsymbol{x}_t}\left[\|s_\theta(\boldsymbol{x}_t, t) - \nabla \log p_{t|0}(\boldsymbol{x}_t|\boldsymbol{x}_0)\|_2^2\right],
\end{equation}
where \(t\sim\mathcal{U}(0,1)\) and \(p_{t|0}\) is the conditional transition distribution from \(\boldsymbol{x}_0\) to \(\boldsymbol{x}_t\), which is determined by the pre-defined forward SDE and is analytical for a linear drift \(\boldsymbol{f}(\boldsymbol{x}_t,t) = f(t)\boldsymbol{x}_t\). 

In this context, the score network \(s_\theta(\boldsymbol{x}_t, t)\) is trained to approximate the gradient of the log probability density \(\nabla \log p_{t|0}(\boldsymbol{x}_t|\boldsymbol{x}_0)\), which represents the score function. By minimizing the MSE between the score network's output and the true score function, the network learns to denoise the perturbed data samples \(\boldsymbol{x}_t\) at different time steps \(t\). This approach leverages the forward SDE to model the data distribution and enables effective sampling from the learned generative model by reversing the diffusion process.

\subsection{\label{subsec:flowmatching} Flow Matching Techniques}
Song et al.~\cite{song:scorebased} prove the existence of an ordinary differential equation (ODE), namely the probability flow ODE, whose trajectories have the same marginals as the reverse time SDE in Eq.~\eqref{eq:revsde}. The probability flow ODE is given by:
\begin{align}
    \mathrm{d}\boldsymbol{x}_t=[\boldsymbol{f}(\boldsymbol{x}_t,t)-\frac{1}{2}g^2(t)\nabla\log p_t(\boldsymbol{x}_t)]\mathrm{d}t,\nonumber \\
    \boldsymbol{x}_1\sim p_1\approx p_\mathrm{prior}.
    \label{eq:revode}
\end{align}

Both the reverse-time SDE and the probability flow ODE allow sampling from the same data distribution as their trajectories have the same marginals.

To bridge the theoretical foundation of probability flow ODEs with modern flow-based approaches, we now introduce the framework of Flow Matching – a unified perspective for constructing generative models through ordinary differential equations~\cite{kingma2013auto}. At its core, Flow Matching aims to learn a time-dependent vector field \( v(t, \mathbf{x}) \) that transports samples from a simple prior distribution \( p_0 \) to the target data distribution \( p_1 \) via the ODE:
\begin{equation}
    \frac{d\mathbf{x}_t}{dt} = v(t, \mathbf{x}_t), \quad \mathbf{x}_0 \sim p_0=p_\mathrm{data}.
\end{equation}

This framework attempts to generalize the probability flow ODE in Eq.~\eqref{eq:revode}, where the drift coefficient \( \boldsymbol{f}(\boldsymbol{x}_t,t) - \frac{1}{2}g^2(t)\nabla\log p_t(\boldsymbol{x}_t) \) can be viewed as a specific instantiation of the flow field \( v(t, \mathbf{x}) \)~\cite{rezende2015variational}. The key innovation lies in learning \( v(t, \mathbf{x}) \) directly from data through different matching objectives~\cite{dinh2017density}. Within this framework, two representative approaches have been developed:

\subsubsection{Rectified Flow Matching}  
Rectified flow matching (RFM)~\cite{lipman2023flow} directly minimizes trajectory discrepancies using a simple quadratic loss:
\begin{equation}
    \mathcal{L}_{\text{RFM}} = \mathbb{E}_{t,\mathbf{x}_0,\mathbf{x}_t}\left[\left\|v(t, \mathbf{x}_t) - \frac{\mathbf{x}_t - \mathbf{x}_0}{t}\right\|_2^2\right],
    \label{eq:rfmloss}
\end{equation}
where \( \mathbf{x}_t \) represents intermediate states. This approach emphasizes direct alignment of the learned flow with linear interpolants between samples.

\subsubsection{Optimal-Transport Flow Matching} 
Chen \emph{et al.}~\cite{chen2018neural} introduced Neural ODEs, which provide a continuous-time generalization of residual networks and have been applied to generative modeling. Optimal-Transport Flow Matching (OTFM) incorporates Wasserstein-2 optimal transport theory~\cite{villani2008optimal} through a potential-guided loss:
\begin{equation}
   \mathcal{L}_{\text{OTFM}} = \mathbb{E}_{t,\mathbf{x}_t}\left[\left\|v(t, \mathbf{x}_t) - \nabla_{\mathbf{x}_t}\phi(t, \mathbf{x}_t)\right\|_2^2\right],
   \label{eq:otfmloss}
\end{equation}
where \( \phi(t,\mathbf{x}) \) is the Kantorovich potential~\cite{kantorovich2006translocation} from the optimal transport theory. This formulation explicitly minimizes the kinetic energy of the transport map by leveraging the geometric properties of Wasserstein space.

While both approaches share the common goal of learning transport dynamics, their fundamental differences manifest in three key aspects: First, in \emph{objective focus}, RFM prioritizes sample trajectory alignment through linear interpolation, whereas OTFM minimizes optimal transport costs via Wasserstein-2 geometry~\cite{ambrosio2008gradient}. Second, their \emph{theoretical foundations} differ substantially – OTFM directly incorporates the differential structure of optimal transport through the Kantorovich potential \( \phi \), while RFM relies on simpler interpolation matching without explicit geometric constraints. Third, there exist notable \emph{computational tradeoffs}: RFM's quadratic loss admits efficient training through closed-form solutions, while OTFM requires solving the dual potential formulation that increases computational complexity.

\section{\label{sec:FNSE-SBGAN} FNSE-SBGAN}
In this section, we introduce the FNSE-SBGAN, an SB-based T-F domain far-field to near-field SE method. We start by defining the SB concept and the paired data for the SE task based on the signal model described in Section~\ref{subsec:Signal-Models}. Next, we provide a comprehensive description of the loss functions employed during training. Finally, we summarize the training and inference process of FNSE-SBGAN.

\subsection{\label{subsec:3:2} Schrödinger Bridge}
The SB problem~\cite{schrodinger:theorie,DeBortoli:940392f5} originates from the optimization of path measures with constrained boundaries. We define the target distribution \(p_{\mathrm{X}}\) to be equal to the data distribution \(p_{\mathrm{data}}\), and we consider the distribution of mixture signals \(M\), denoted as \(p_\mathrm{M}\), to be the prior distribution. Considering \(p_0,p_1\) as the marginal distributions of \(p\) at boundaries, SB is defined as minimization of the Kullback-Leibler (KL) divergence:
\begin{equation}
    \min_{p\in\mathcal{P}_{[0,1]}}D_{\mathrm{KL}}(p\parallel p_{\mathrm{ref}}),\quad s.t. \ p_0=p_{\mathrm{X}},\ p_1=p_\mathrm{M}, 
\end{equation}
where \(\mathcal{P}_{[0,1]}\) is the space of path measures on a finite time interval \([0,1]\) with \(p_{\mathrm{ref}}\) denoting the reference path measure. When \(p_{\mathrm{ref}}\) is defined by the same form of forward SDE as SGMs in Eq. (\ref{eq:sde}), the SB problem is equivalent to a pair of forward-backward SDEs~\cite{wang:pmlr,chen:likelihood}:
\begin{small}
    \begin{equation}
        \mathrm{d}\boldsymbol{x}_t=[\boldsymbol{f}(\boldsymbol{x}_t,t)+g^2(t)\nabla\log\Psi_t(\boldsymbol{x}_t)]\mathrm{d}t+g(t)\mathrm{d}\boldsymbol{w}_t,\quad\boldsymbol{x}_0\sim p_\mathrm{X},
        \label{eq:couple1}
    \end{equation}
    \begin{equation}
        \mathrm{d}\boldsymbol{x}_t=[\boldsymbol{f}(\boldsymbol{x}_t,t)-g^2(t)\nabla\log\widehat{\Psi}_t(\boldsymbol{x}_t)]\mathrm{d}t+g(t)\mathrm{d}\bar{\boldsymbol{w}}_t,\quad\boldsymbol{x}_1\sim p_{\mathrm{M}},
        \label{eq:couple2}
    \end{equation}
\end{small}where \(\boldsymbol{f}\) , \(g\) and \(\boldsymbol{w}_t\) are from the reference SDE in Eq. (\ref{eq:sde}). With \(\Psi_t\) and \(\widehat{\Psi}_t\) representing the optimal forward and reverse drifts, the marginal distribution of the SB state \(\boldsymbol{x}_t\) can be expressed as \(p_t=\widehat{\Psi}_t\Psi_t\). Typically, SB is not fully tractable - closed-form solutions exist only when the families of \(p_{\mathrm{ref}}\) are strictly limited~\cite{Bunne:pmlr-v206-bunne23a,chen:schrodinger}.

\subsection{\label{subsec:3:3} Schrödinger Bridge between Paired Data}
Exploring the tractable SB between Gaussian-smoothed paired data with linear drift in SDE, we consider Gaussian boundary conditions \(p_\mathrm{X} = \mathcal{N}_\mathbb{C}(\boldsymbol{x}_0, \epsilon_0^2 \mathbf{I})\) and \(p_{\mathrm{M}} = \mathcal{N}_\mathbb{C}(\boldsymbol{x}_{1},e^{2\int_{0}^{1}f(\tau)\mathrm{d}\tau}\epsilon_0^2\boldsymbol{I})\). As \(\epsilon_0 \to 0\), \(\widehat{\Psi}_t\) and \(\Psi_t\) converge to the tractable solutions between the target data \(\boldsymbol{x}_0\) and the corrupted data \(\boldsymbol{x}_1\):
\begin{align}
\widehat{\Psi}_t=\mathcal{N}_\mathbb{C}(\alpha_t\boldsymbol{x}_0,\alpha_t^2\sigma_t^2\boldsymbol{I}),
\Psi_t=\mathcal{N}_\mathbb{C}(\bar{\alpha}_t\boldsymbol{x}_1,\alpha_t^2\bar{\sigma}_t^2\boldsymbol{I}),
\label{eq:couple_psi}
\end{align}
where \(\alpha_t=e^{\int_0^tf(\tau)\mathrm{d}\tau}\), \(\bar{\alpha}_t=e^{-\int_t^1f(\tau)\mathrm{d}\tau}\), \(\sigma_t^2=\int_0^t\frac{g^2(\tau)}{\alpha_\tau^2}\mathrm{d}\tau\)  and \(\bar{\sigma}_t^2=\int_t^1\frac{g^2(\tau)}{\alpha_\tau^2}\mathrm{d}\tau\) are determined by  \(f\) and \(g\) in the reference SDE, which are analogous to the noise schedule in SGMs~\cite{kingma:variational}. The marginal distribution of SB also has a tractable form: 
\begin{equation}
p_t=\Psi_t\widehat{\Psi}_t=\mathcal{N}\left(\frac{\alpha_t\bar{\sigma}_t^2\boldsymbol{x}_0+\bar{\alpha}_t\sigma_t^2\boldsymbol{x}_1}{\sigma_1^2},\frac{\alpha_t^2\bar{\sigma}_t^2\sigma_t^2}{\sigma_1^2}\boldsymbol{I}\right).
\label{eq:marginal-distribution}
\end{equation}

Several noise schedules~\cite{chen:schrodinger,jukic:SBSE}, such as variance-preserving (VP), variance-exploding (VE) and gmax, are listed in Table~\ref{tab:noiseschedules}. 

\begin{table}
    \centering
    \caption{Demonstration of the noise schedules in FNSE-SBGAN. "Sch." denotes the schedule. }
    \resizebox{0.48\textwidth}{!}{
    \begin{tabular}{c|ccc}
        \hline
        \multicolumn{1}{c}{Sch.} & \multicolumn{1}{c}{gmax} & \multicolumn{1}{c}{Scaled VP} & \multicolumn{1}{c}{VE} \\
        \hline
        \(f(t)\)      & 0            & \(-\frac{1}{2}(\beta_0+t(\beta_1-\beta_0))\)& 0  \\
        \(g^2(t)\)    & \(\beta_0+t(\beta_1-\beta_0)\)& \(c(\beta_0+t(\beta_1-\beta_0))\)&\(ck^{2t}\) \\
        \(\alpha_t\)  & 1            & \(e^{-\frac{1}{2}\int_0^t(\beta_0+\tau(\beta_1-\beta_0))\mathrm{d}\tau}\)& 1  \\
        \(\sigma_t^2\)& \(\frac{t^2(\beta_1-\beta_0)}{2}+\beta_0t\)& \(c(e^{\int_0^t(\beta_0+\tau(\beta_1-\beta_0))\mathrm{d}\tau}-1)\)& \(\frac{c\left(k^{2t}-1\right)}{2\log(k)}\) \\
        \hline
    \end{tabular}}

\label{tab:noiseschedules}
\end{table}

\subsection{\label{subsec:3:4} Loss Functions}
Following the approach in~\cite{jukic:SBSE}, we train the neural model \(B_{\theta}\) to directly predict the target data, forming the bridge loss~\cite{chen:schrodinger} in Eq.~(\ref{eq:mseloss}). We further augment the bridge loss with a series of reconstruction losses, including the mel loss and phase loss. Additionally, we empirically observe that introducing adversarial loss can effectively improve the generation quality. The overall architecture of our proposed method integrating SB and the adversarial training is demonstrated in Figure~\ref{fig:overview}(f), alongside with other baseline methods described in Sec.~\ref{subsec:4:3}. 

The reconstruction loss includes both the mean-square error (MSE) loss $\mathcal{L}_{mse}$, the mel loss $\mathcal{L}_{mel}$ and the the phase loss $\mathcal{L}_{p}$ following the settings in~{\cite{ai:apnet,du:apnet2}}.  
Let \(X\) denote the target signal and let \(\tilde{X} = B_{\theta}(\boldsymbol{x}_t, \boldsymbol{x}_1, t)\) represent the current estimate produced by the neural network $B_{\theta}$. The first term is defined as the MSE between \(\tilde{X}\) and \(X\) in the STFT domain:
\begin{equation}
    \mathcal{L}_{mse} = \frac{1}{LF} \sum_{l,f} \left\| \tilde{X}_{l,f} - X_{l,f} \right\|_2^2.
    \label{eq:mseloss}
\end{equation}

The mel loss measures the mean absolute error (MAE) between the mel-spectrograms of the estimated \(\tilde{Y}^{mel}\) and target waveforms \(Y^{mel}\):
\begin{equation}
    \mathcal{L}_{mel} = \frac{1}{LF_{mel}} \sum_{l,f} \left\| \tilde{Y}_{l,f}^{mel} - Y_{l,f}^{mel} \right\|_1,
    \label{eq:melloss}
\end{equation}
where the mel-spectrograms \( \{\tilde{Y}_{l,f}^{mel},Y^{mel}\} \in \mathbb{R}^{L \times F_{mel}} \) are obtained from $Y^{mel}=|X|\mathcal{A}$ and $\tilde{Y}^{mel}=|\tilde{X}|\mathcal{A}$, with $\tilde{X}$ the estimated near-field signal and \( \mathcal{A} \in \mathbb{R}^{F \times F_{mel}} \) the linear mel filter. \( F_{mel} \) is the mel size and typically satisfies \( F_{mel} \ll F \) for a compressed representation. 

For the multi-mel loss, we compute the sum of mel losses under the mel filterbank with different scales :
\begin{equation}
    \mathcal{L}_{mel} = \sum_{j=1}^{J}\mathcal{L}_{mel}^{\left(j\right)},
    \label{eq:multimelloss}
\end{equation}
where $J$ denotes the number of scales, and $\mathcal{L}_{mel}^{\left(j\right)}$ refers to the corresponding $j$-th scale mel loss. Following~\cite{kumar2023high}, here $J$ is empirically set to 7. 

The phase loss $\mathcal{L}_{p}$ is usually tricky to optimize due to the wrapping effect. In ~\cite{ai2024apcodec}, a specially designed anti-wrapping phase loss is proposed by incorporating instantaneous phase (IP), group delay (GD) and instantaneous frequency (IF) ~\cite{phase2015Gerkmann}, given by
\begin{equation}
    \mathcal{L}_{p}=\mathcal{L}_{IP}+\mathcal{L}_{GD}+\mathcal{L}_{IF}, 
\end{equation}
where
\begin{equation}
    \mathcal{L}_{IP}=\frac{1}{FL}\sum_{l,f}\left\|f_{AW}\left(\tilde{\Phi}_{l,f}-\Phi_{l,f}\right)\right\|_{1},
\end{equation}
\begin{equation}
    \mathcal{L}_{GD}=\frac{1}{FL}\sum_{l,f}\left\|f_{AW}\left(\Delta_{F}\tilde{\Phi}_{l,f}-\Delta_{F}\Phi_{l,f}\right)\right\|_{1},
\end{equation}
\begin{equation}
    \mathcal{L}_{IF}=\frac{1}{FL}\sum_{l,f}\left\|f_{AW}\left(\Delta_{L}\tilde{\Phi}_{l,f}-\Delta_{L}\Phi_{l,f}\right)\right\|_{1},
\end{equation}
with $\{\Phi,\tilde{\Phi}\} \in \mathbb{C}^{L \times F}$ denoting the near-field and estimated near-field phase spectrum, and $f_{AW}\left(x\right) = \left|x-2\pi\mathrm{round}\left(\frac{x}{2\pi}\right)\right|$ denoting the anti-wrapping loss. $\{\Delta_{F},\Delta_{L}\}$ refer to the differential along the frequency and time axes, respectively.

To further improve the speech quality, the adversarial loss is incorporated. This loss incudes the hinge GANs framework, consisting of discriminators \(D_\phi^{(q)}\) with parameters $\phi$ and a generator. The backpropagated losses are denoted as \(\mathcal{L}_d\) and \(\mathcal{L}_g\), respectively:
\begin{equation}
    \mathcal{L}_d=\frac{1}{Q}\sum_{q=1}^Q\max\left(0,1-D_\phi^{(q)}\left(\mathbf{x}\right)\right)+\max\left(0,1+D_\phi^{(q)}\left(\mathbf{\tilde{x}}\right)\right),
    \label{eq:GAN0}
\end{equation}
\begin{equation}
    \mathcal{L}_g=\frac{1}{Q}\sum_{q=1}^Q\max\left(0,1-D_\phi^{(q)}\left(\mathbf{\tilde{x}}\right)\right),
\end{equation}
where $\tilde{\mathbf{x}}=\text{iSTFT}(\tilde{\mathbf{X}})$ and $\mathbf{x} $ denote the reconstructed waveform and the cleaned near-field waveform, respectively. $\text{iSTFT}\left(\cdot\right)$ refers to the iSTFT operation,
and \(Q\) is the number of sub-discriminators. Discriminators include multi-period discriminator (MPD)~\cite{kong:hifigan} and multi-resolution spectrogram discriminator (MRSD)~\cite{Jang:UnivNetAN}. 

Besides, the feature matching loss is also utilized:
\begin{equation}
    \mathcal{L}_{fm}=\frac{1}{IQ}\sum_{i,q}|\mathbf{f}_i^{(q)}\left(\mathbf{\tilde{x}}\right)-\mathbf{f}_i^{(q)}\left(\mathbf{x}\right)|,
\end{equation}
where \(\mathbf{f}_i^{(q)}(\cdot)\) denotes the \(i\)-th layer feature for the \(q\)-th sub-discriminator. Finally, the loss for the neural model is formulated as
\begin{equation}
    \mathcal{L}_{B}=\mathcal{L}_{mse}+\lambda_{mel}\mathcal{L}_{mel}+\lambda_p\mathcal{L}_{p}+\lambda_g\mathcal{L}_g+\lambda_{fm}\mathcal{L}_{fm},
    \label{eq:loss}
\end{equation}
where \(\lambda_{mel}\), $\lambda_{p}$, \(\lambda_{g}\) and \(\lambda_{fm}\) are the weight hyperparameters of the corresponding loss.

\subsection{Algorithms}
\begin{algorithm}[ht]
\caption{FNSE-SBGAN Training (Based on SDE).}
\begin{algorithmic}
\STATE 
\STATE {\textsc{Input}} Training set of pairs $(\mathbf{M}, \mathbf{X})$, labeled as $(\boldsymbol{x}_1, \boldsymbol{x}_0)$
\STATE {\textsc{Output}} Trained parameters $\{\theta, \phi\}$
\STATE 1: Sample diffusion time $t\sim\mathcal{U}(t_{min},1)$
\STATE 2: Generate perturbed state $x_t$ using the marginal distribution in Eq.~(\ref{eq:marginal-distribution})
\STATE 3: Estimate $\tilde{X} \gets \tilde{\boldsymbol{x}}_0 = B_{\theta}(\boldsymbol{x}_t, \boldsymbol{x}_1, t)$
\STATE 4: Estimate the discriminative labels with \(D_\phi\)
\STATE 5: Compute the discriminator loss $\mathcal{L}_d$ with Eq.~(\ref{eq:GAN0}) 
\STATE 6: Compute the generator loss $\mathcal{L}_{B}$ with Eq.~(\ref{eq:loss})
\STATE 7: Backpropagate loss $\mathcal{L}_d$ to update $\phi$
\STATE 8: Backpropagate loss $\mathcal{L}_{B}$ to update $\theta$
\end{algorithmic}
\label{alg:FNSE-SBGAN-Training}
\end{algorithm}

\begin{algorithm}[ht]
\caption{FNSE-SBGAN Inference (Based on SDE).}
\begin{algorithmic}
\STATE
\STATE {\textsc{Input}} Far-field speech $\boldsymbol{x}_1\gets \mathbf{M}$, step size $\Delta\tau=\frac{(1-t_{min})}{N}$
\STATE {\textsc{Output}} Near-field speech estimate $\tilde{X}$
\STATE 1: Generate initial reverse state $\boldsymbol{x}_t\gets\boldsymbol{x}_1$
\STATE 2: \textbf{for}  $n \in \{N,...,1\}$ \textbf{do}
\STATE 3: \hspace{1em}Get diffusion time $t=n\Delta\tau=\frac{n}{N}(1-t_{min})$
\STATE 4: \hspace{1em}Compute $\alpha_t$, $\sigma_t^2$, $\alpha_{t-1}$, $\sigma_{t-1}^2$ via Table~\ref{tab:noiseschedules}
\STATE 5: \hspace{1em}Estimate $\tilde{\boldsymbol{x}}_0 = B_\theta(\boldsymbol{x}_t, \boldsymbol{x}_1, t)$
\STATE 6: \hspace{1em}Compute $\boldsymbol{x}_{t-1} = \frac{\alpha_{t-1}}{\alpha_t}\frac{\sigma_{t-1}^2}{\sigma_t^2}\boldsymbol{x}_t + \alpha_{t-1}\left(1-\frac{\sigma_{t-1}^2}{\sigma_{t}^2}\right)\tilde{\boldsymbol{x}}_0$
\STATE 7: \hspace{1em}Sample noise signal $\boldsymbol{\epsilon}\sim\mathcal{N}(0,\mathbf{I})$
\STATE 8: \hspace{1em}\textbf{if} $n \neq 1$
\STATE  \hspace{3em} update $\boldsymbol{x}_{t-1} \gets \boldsymbol{x}_{t-1} + \alpha_{t-1}\sqrt{\sigma_{t-1}^2\left(1-\frac{\sigma_{t-1}^2}{\sigma_t^2}\right)}\boldsymbol{\epsilon}$
\STATE 9: \textbf{end for}
\STATE10: Return $\tilde{X}\gets \boldsymbol{x}_0$
\end{algorithmic}
\label{alg:FNSE-SBGAN-Infer}
\end{algorithm}


\section{\label{sec:result} Results and Discussions}
\subsection{\label{subsec:datasets} Datasets}
In our study, we utilize the same Chinese real-recorded datasets, with detailed data collection procedures, feature analysis, and preprocessing methods as described in ~\cite{LEI:FNSESAT}. However, we observe that augmenting far-field signals by adding extra noise, which is effective for FNSE-SAT, has no significant effect on our SB-based diffusion method and, in some cases, even negatively impacts performance. Therefore, in this work, we list the performances of different methods on the original 400 hours of recorded signal pairs and the expanded 1000 hours of signal pairs. When the Data++ option is set to \ding{55}, it indicates that the original 400 hours of recorded signal pairs are used for training. When the Data++ option is set to $\checkmark$, our training dataset is expanded to 1,000 hours by adding extra noise from the DNS Challenge 3 dataset \footnote{https://github.com/microsoft/DNS-Challenge}. Our test set consists of 497 far- and near-field pairs, ensuring no overlap with the training set. 

\subsection{\label{subsec:4:2} Experimental configurations}
In terms of noise schedulers, \(\beta_0 = 0.01\) and \(\beta_1 = 20\) are set for both gmax and scaled variance preservation (VP) types; For variance exploding (VE) type, we use \(k=2.6\) and \(c=0.40\); For scaled VP type, we use \(c=0.30\). The processing time for the proposed SB is set to \(T=1\) with $t_{\text{min}} = 10^{-4}$. The reverse SDE and the probability flow ODE~\cite{chen:likelihood} samplers are chosen in the inference stage by changing
 \(p_{\mathrm{ref}}\) of Eq.~(\ref{eq:sde}) to Eq.~(\ref{eq:revode}). For flow matching techniques, the Euler sampler~\cite{anderson1982reverse} is applied.

For the weight hyperparameters in Eq.~(\ref{eq:loss}), \(\lambda_{mel}\), $\lambda_{p}$, \(\lambda_{g}\) and \(\lambda_{fm}\) are 0.1, 0.01, 10.0 and 10.0, respectively. ``+GAN" refers to the inclusion of the loss terms \(\mathcal{L}_g\) and \(\mathcal{L}_{fm}\) in Eq.~(\ref{eq:loss}). For feature extraction, we employ STFT with a Hanning window of length 1024 and a hop size of 256. All models are non-causal. The parameter size of TF-GridNet\cite{wang2023tf} in FNSE-SAT remains the same as in \cite{LEI:FNSESAT}, but the stride of the Unfold and Deconv1D layers changes from 4 to 1, resulting in higher computational complexity.

The configurations in Eq.~(\ref{eq:multimelloss}) vary in the FFT size (\(n_{\text{fft}}\)) and the number of mel frequency bins (\(n_{\text{mels}}\)), which are set to (32, 64, 128, 256, 512, 1024, 2048) and (5, 10, 20, 40, 80, 160, 210), respectively. For all configurations, the upper-bound frequency (\(f_{\text{max}}\)) is fixed at half the sampling rate, while the window size and hop size are set to \(n_{\text{fft}}\) and \(n_{\text{fft}}/4\), respectively. The single mel loss defined in Eq.~(\ref{eq:melloss}) is acquired with the specific configuration where \(n_{\text{fft}} = 1024\) and \(n_{\text{mels}} = 160\).

The discriminator in Eq.~(\ref{eq:GAN0}) includes two components: Multi-Period Discriminator (MPD) and Multi-Resolution Spectral Discriminator (MRSD). MPD captures variations in audio periodic patterns using five sub-discriminators. Each sub-discriminator reshapes the 1D raw audio waveform into a 2D format based on predefined period values, which are set to \{2, 3, 5, 7, 11\}. MRSD consists of three sub-discriminators, where the magnitude spectrum serves as the input. This input is fed into a stack of Conv2d layers to compute the discriminative scores. The configurations for \{window size, hop size, \(n_{\text{fft}}\)\} in the three sub-discriminators are (512, 128, 512), (1024, 256, 1024), and (2048, 512, 2048), respectively.

The NCSN++ network with 16.2 M parameters~\cite{song:scorebased} is modified based on the default network configuration used in SGMSE~\cite{sgmse} through several adjustments. The embedding size is set to 64, the channel multiplier is set to (1, 1, 2, 2, 2, 2, 2), the number of residual blocks is increased to 2, and the attention resolution is removed. For the configurations with 36.5 M and 64.9 M parameters, the embedding size is adjusted to 96 and 128, respectively.

\begin{figure}[!t]
    \centering
    \includegraphics[width=1.0\linewidth]{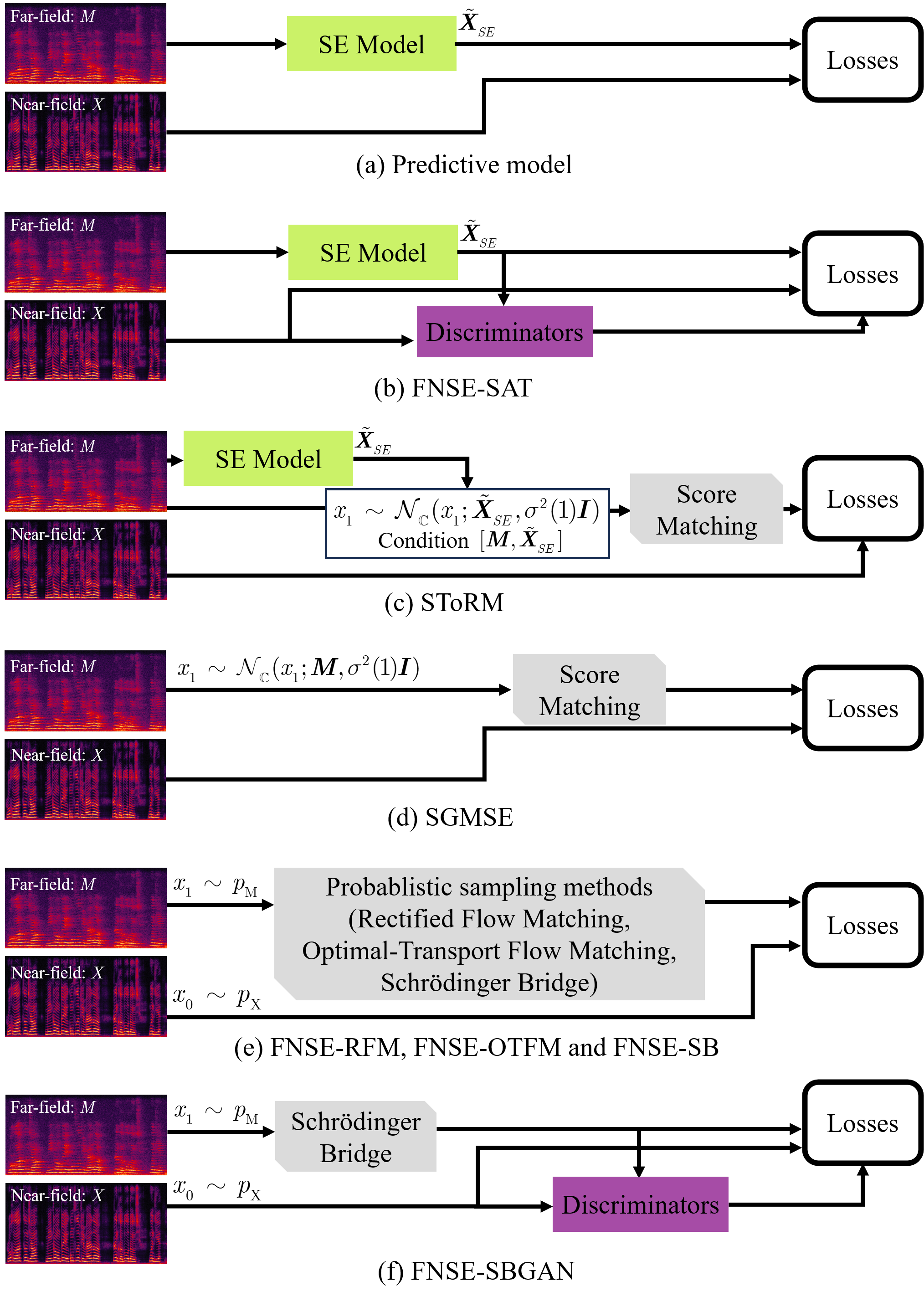}
    \caption{Illustrations of several baselines and our proposed FNSE-SBGAN.}
    \label{fig:overview}
\end{figure}

\subsection{\label{subsec:baselines} Baselines Comparisons}

Due to the relatively limited research in this field, we select several state-of-the-art (SOTA) SE algorithms with different paradigms as baselines, shown in Figure~\ref{fig:overview}(a)-(e), with detailed descriptions below.

\begin{itemize}
    \item As shown in Figure~\ref{fig:overview}(a), ``Predictive model" indicates that no generative strategy is used, and the model directly predicts the target in an end-to-end manner. Here, we use TF-GridNet~\cite{wang2023tf} as the backbone.
    \item ``FNSE-SAT" refers to using TF-GridNet as the generator and assisting it with a GAN structure to generate near-field speech, as illustrated in Figure~\ref{fig:overview}(b).
    \item ``SToRM" represents the two-stage method in~\cite{lemercier2023storm}, where the result estimated by the predictive model (NCSN++~\cite{song:scorebased}) in the first stage is concatenated with the far-field signal to serve as the condition for diffusion in the second stage. The estimated signal from the first stage, perturbed by white noise, serves as the starting point for the score matching (NCSN++) diffusion process with a one-step corrector, as shown in Figure~\ref{fig:overview}(c). 
    \item ``SGMSE" from~\cite{sgmse} is a score matching method with an OUVE and a one-step corrector~\cite{song:scorebased}, shown in Figure~\ref{fig:overview}(d).
    \item The variations of ``FNSE-SB" with several different probabilistic sampling methods are shown in Figure~\ref{fig:overview}(e), using rectified flow matching~\cite{guo2024voiceflow} and optimal-transport flow matching~\cite{mehta2024matcha}, within the same training framework. Both flow matching methods employ an Euler sampler for the ODE. Unlike the score network $s_\theta$, which takes \(\nabla\log{p_{t}}\) as its training objective, or the flow matching network ${O}_\theta$ that predicts the velocity field to transform the initial distribution into the target distribution, the SB network ${B}_\theta$ directly predicts the target data. Here, $s_\theta$, ${O}_\theta$ and ${B}_\theta$ represent the same NCSN++ network, but they estimate different objectives based on the sampling methods of score matching, flow matching, and SB, respectively. ``FNSE-RFM", ``FNSE-OTFM" and ``FNSE-SB" correspond to rectified flow matching, optimal-transport flow matching and Schrödinger Bridge without GAN loss.
    \item Finally, our proposed FNSE-SBGAN model is illustrated in Figure~\ref{fig:overview}(f).
\end{itemize}

\begin{figure}[!t]
    \centering
    \includegraphics[width=1.0\linewidth]{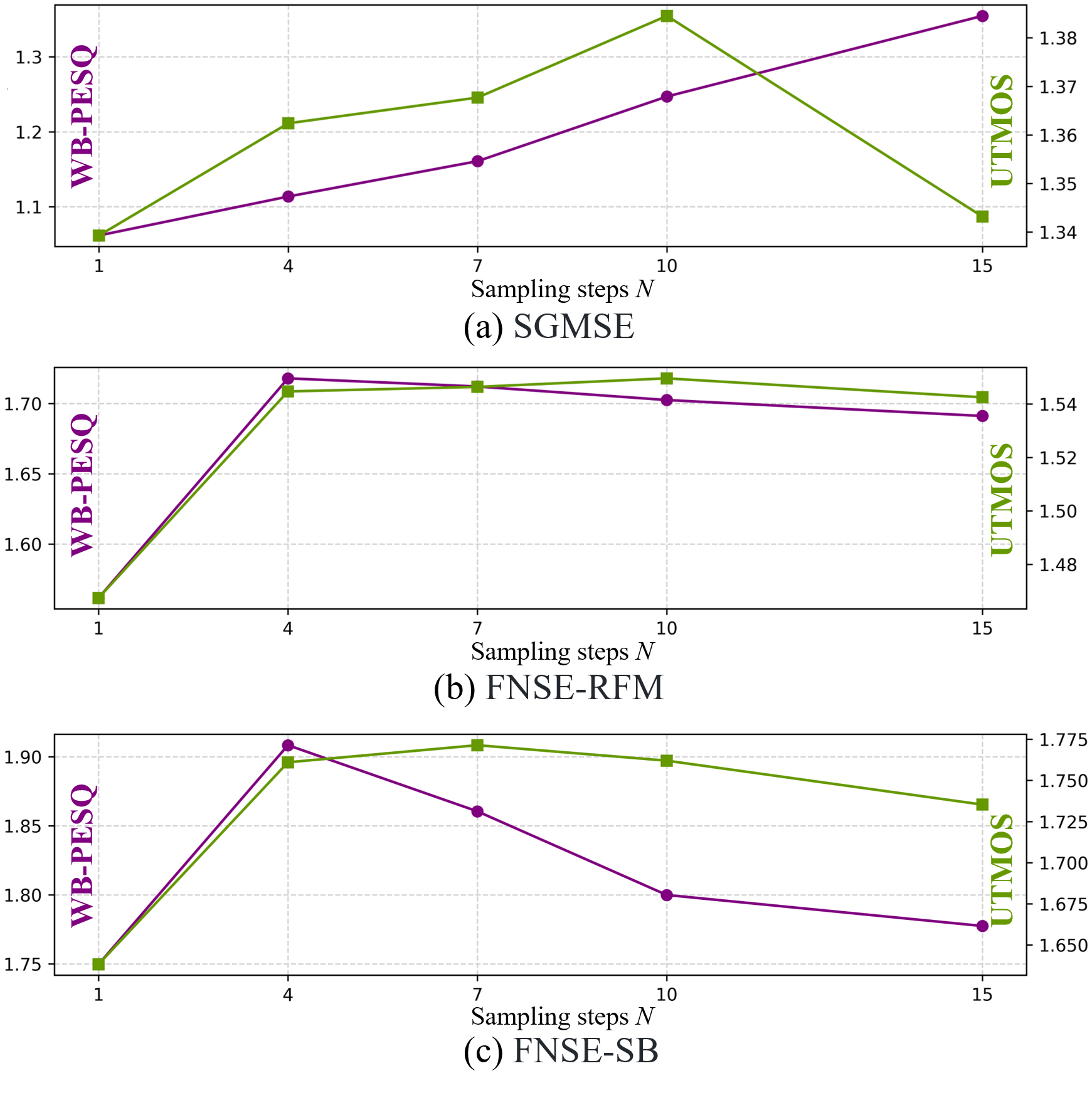}
    \vspace{-16PT}
    \caption{Metrics with different numbers of sampling steps $N$ during the reverse process.}
    \label{fig:reverseN}
\end{figure}

\subsection{\label{subsec:4:3} Results and Analysis}
Five metrics are involved in the objective evaluations: 1) Wide-band version of Perceptual evaluation of speech quality (PESQ)~{\cite{rec2005p}} serves to assess the objective speech quality. 2) Extended Short-Time Objective Intelligibility (ESTOI)~{\cite{taal2011algorithm}} measures the intelligibility level of speech. 3) UTMOS~{\cite{saeki2022utmos}} is used to obtain subjective scores related to the perceived quality of speech, providing an objective approximation of human judgment. 4) DNSMOS P.808 is a non-intrusive metric to provide a comprehensive assessment of the model performance ~\cite{reddy2021dnsmos}. 5) The Character Error Rate (CER), computed between the reference text and the transcription generated by Whisper\footnote{https://github.com/openai/whisper} on the enhanced signal, is used to assess the degree of semantic modification introduced by the generation method. 

\renewcommand\arraystretch{1.03}
\begin{table}[t]
	\centering
        \caption{Ablation Study of Loss Functions, Noise Schedules, and Samplers in Signal Reconstruction Using Mapping Method with 16.2 M Parameters, Cleaned Near-field targets, and Exclusion of Data++.}
	\Huge
	\resizebox{0.48\textwidth}{!}{
    \begin{tabular}{ccc|ccc}
        \hline
			Schedules&Losses &Sampler&WB-PESQ\(\uparrow\)& ESTOI\(\uparrow\)&UTMOS\(\uparrow\)\\
        \hline
        gmax&mse &SDE&1.714& 0.6861& 1.594\\
        gmax&+mmel &SDE&1.908&    0.7409& 1.761\\
 gmax&  +mmel &ODE& 1.846& 0.7337&1.674\\
        Scaled VP 
&+mmel &SDE&1.882&  0.7401& 1.698\\
 VE& +mmel &SDE&1.879& 0.7398& 1.665\\
 gmax& +mmel+phase &SDE&1.915& 0.7482&1.785\\
 gmax& +mmel+GAN &SDE&1.937& 0.7499&1.789\\
        gmax&+mmel+phase+GAN &SDE&1.941& 0.7501& 1.790\\
        \hline
    \end{tabular}}
    \label{tab:abla-sched-loss-result}
\end{table}

\begin{table}
    \centering
    \caption{Ablation Study of Signal Reconstruction Methods and Network Sizes Under the Conditions of ``gmax", ``SDE", and ``mse+mmel" Losses.}
    \Huge
    \resizebox{0.45\textwidth}{!}{
    \begin{tabular}{cc|cc|ccc}
        \hline
        \multirow{2}*{Recon.}&   \#Param. &Data &Cleaned&  \multirow{2}*{WB-PESQ\(\uparrow\)}& \multirow{2}*{ESTOI\(\uparrow\)}&  \multirow{2}*{UTMOS\(\uparrow\)}\\
          &  (M) & ++ & Near-field & &&\\
        \hline
        map& 16.2 &\ding{55}&$\checkmark$& 1.908& 0.7409&  1.761\\
 map& 16.2 & \ding{55}& \ding{55}& 1.645& 0.7172&1.648\\
        crm& 16.2 &$\checkmark$&$\checkmark$& 1.875& 0.7441& 1.735\\
 crm& 16.2 & \ding{55}& $\checkmark$& 1.943& 0.7532&1.780\\
        decouple& 16.2 & \ding{55}&$\checkmark$& 1.931& 0.7507& 1.775\\
        crm& 36.5 & \ding{55}&$\checkmark$& 1.972& 0.8204& 1.839\\
        crm& 64.9 & \ding{55}&$\checkmark$& 2.023& 0.8642& 1.879\\
        \hline
    \end{tabular}}
    \label{tab:abla-recon-net-result}
\end{table}

\begin{table*}
    \centering
    \caption{Results of objective evaluations on the test set. ``\#Param." denotes the number of trainable parameters. Metrics with \(\downarrow\) indicate that lower values are better. The inference speed on a GPU is evaluated based on a single Tesla V100. The computational complexity of the diffusion methods needs to be multiplied $\times$ by the number of reverse sampling steps. The SToRM method combines predictive and diffusion models, so the number of parameters and computational complexity need to be summed + from both parts. The best and second-best performances are namely highlighted in \textbf{bold} and \underline{underlined}.}
    \vspace{-8pt}
    \Huge
    \resizebox{1.0\textwidth}{!}{
    \begin{tabular}{llll|c|ccccc}
        \hline
	\multirow{2}*{Models} & \#Param.  & \multirow{2}*{\#MACs (Giga/5s) } & Inference  &\multirow{2}*{Data++}&\multirow{2}*{WB-PESQ\(\uparrow\)}& \multirow{2}*{ESTOI\(\uparrow\)} & \multirow{2}*{UTMOS\(\uparrow\)} & \multirow{2}*{DNSMOS\(\uparrow\)}& CER   \\
         & (M) &   & Speed &  &  &  &    &   & (in \%) $\downarrow$ \\
        \hline
        Real-near& -& -& - & -& 4.644& 1.000& 2.989& 3.915&8.921\\
        Real-far& -& -& - &-&1.125&  0.3772& 1.286& 2.451& 35.94\\
        \hline
        Predictive& 2.52& 182.3& 0.0280 & $\checkmark$& 1.609& 0.6485& 1.351& 3.135&27.11\\
        FNSE-SAT& 2.52& 182.3& 0.0280 &\ding{55}& 1.637& 0.6513& 1.473& 3.341&26.87\\
        FNSE-SAT& 2.52& 182.3& 0.0280 &$\checkmark$&1.814&  0.7083& 1.781& 3.667& 26.09\\
        SToRM& 16.2+16.2& 82.57+82.80$\times$20& 0.2669&$\checkmark$&1.586&  0.6368& 1.590& 3.393& 44.98\\
        FNSE-SAT-SToRM& 2.52+16.2& 182.3+82.80$\times$20& 0.2823&$\checkmark$&1.905&  0.7379& \underline{1.840}& \underline{3.795}& 27.76\\
        SGMSE& 16.2  & 82.57$\times$20& 0.2535&\ding{55}& 1.247&  0.5680& 1.385& 3.375& 51.01\\
        \hline
        FNSE-RFM (ours)& 16.2  & 82.57$\times$4& 0.0506&\ding{55}& 1.718&  0.6901& 1.545& 3.469& 36.49\\
        FNSE-OTFM  (ours)& 16.2  & 82.57$\times$4& 0.0506&\ding{55}& 1.710&  0.6875& 1.537& 3.477& 36.77\\
 FNSE-SB  (ours)& 16.2  & 82.57$\times$4& 0.0507& \ding{55}& 1.950& 0.7564& 1.805& 3.755&24.99\\
        FNSE-SBGAN  (ours)& 16.2  & 82.57$\times$4& 0.0507&\ding{55}& \underline{1.981}&  \underline{0.7582}& 1.812& 3.767& \underline{24.42}\\
        FNSE-SBGAN-Large (ours)&  64.8  & 326.9$\times$4&0.1570&\ding{55}&\textbf{2.131}& \textbf{0.7802}& \textbf{1.992}& \textbf{3.796}& \textbf{21.36}\\
        \hline
    \end{tabular}}
    \label{tab:baseline-result}
\end{table*}

\subsubsection{Ablation studies}

Figure~\ref{fig:reverseN} shows the results of the ablation study on the number of reverse sampling steps for several probabilistic sampling methods. The settings for ``FNSE-SB" are ``gmax" / ``mse+mmel" / ``map" / ``16.2M". The settings for ``SGMSE" and ``FNSE-RFM" are ``mse" / ``map" / ``16.2M". We found that ``SGMSE" requires about 10 steps to achieve the optimal effect. Since a one-step corrector is needed to calibrate the estimate of \(\nabla\log{p_{t}}\), the 10-step iteration process requires passing through the score network \(s_\theta(\boldsymbol{x}_t,t)\) 20 times, which is computationally intensive. In contrast, ``FNSE-SB" and ``FNSE-RFM" only need about four steps to achieve the optimal effect and do not require a corrector, significantly improving efficiency.

Table~\ref{tab:abla-sched-loss-result} and Table~\ref{tab:abla-recon-net-result} reveal the process of exploring the optimal settings for "FNSE-SBGAN". Table~\ref{tab:abla-sched-loss-result} presents the test performance with various combinations of losses and noise schedules when the network parameter count is 16.2M. From the experimental results, it is evident that the introduction of auxiliary losses, single mel loss ``+mel", multi-mel loss ``+mmel", and phase loss ``+phase", can significantly enhance the model's performance. Furthermore, the introduction of adversarial training on top of ``+mmel" and ``+mmel+phase" further improve the WB-PESQ score by 0.091 and 0.026 respectively. Correspondingly, other metrics also show notable improvements. When comparing Scaled VP and VE under the ``+mmel" condition, gmax emerges as the optimal choice for the majority of indicators. Additionally, when the sampler is switched from the reverse SDE to the probability flow ODE, there is a slight degradation in performance.

Table~\ref{tab:abla-recon-net-result} lists the results for the methods of reconstructing the signal from the network output and varying the network size under the settings of ``gmax", ``+mmel", and ``SDE". ``map" and ``crm" denote that the network output is the complex spectrum mapping and the complex mask, respectively. ``decouple" indicates that the network outputs the amplitude and phase of the signal separately, which are then coupled to form the output signal. The results indicate that the ``crm" configuration is optimal for our task, rather than the previously default ``map" form used in the NCSN++ network~\cite{song:scorebased}. Additionally, increasing the network size also improves the final output scores.

Besides, we find that using raw near-field signals as training targets significantly degrades the model's performance due to residual noise and slight reverberation. As we describe in Section~\ref{subsec:datasets}, the data augmentation method of adding extra noise to the far-field signals has a negative impact on our SB-based diffusion method. This is likely because the SDE-based method inherently involves adding Gaussian white noise during the training process. Thus, introducing additional noise may disrupt the intrinsic structure and distribution of the data, causing confusion in the learning process and making it difficult for the model to accurately capture the key information of the signals, thereby affecting the final reconstruction performance~\cite{ho2020denoising}. For instance, as mentioned in~\cite{song2019generative}, the introduction of noise needs to be handled carefully, especially during data augmentation, as excessive noise may obscure the true data distribution, thereby impairing the model's training effectiveness and generation quality.

\begin{figure*}[!t]
        \centering
        \includegraphics[width=1.0\linewidth]{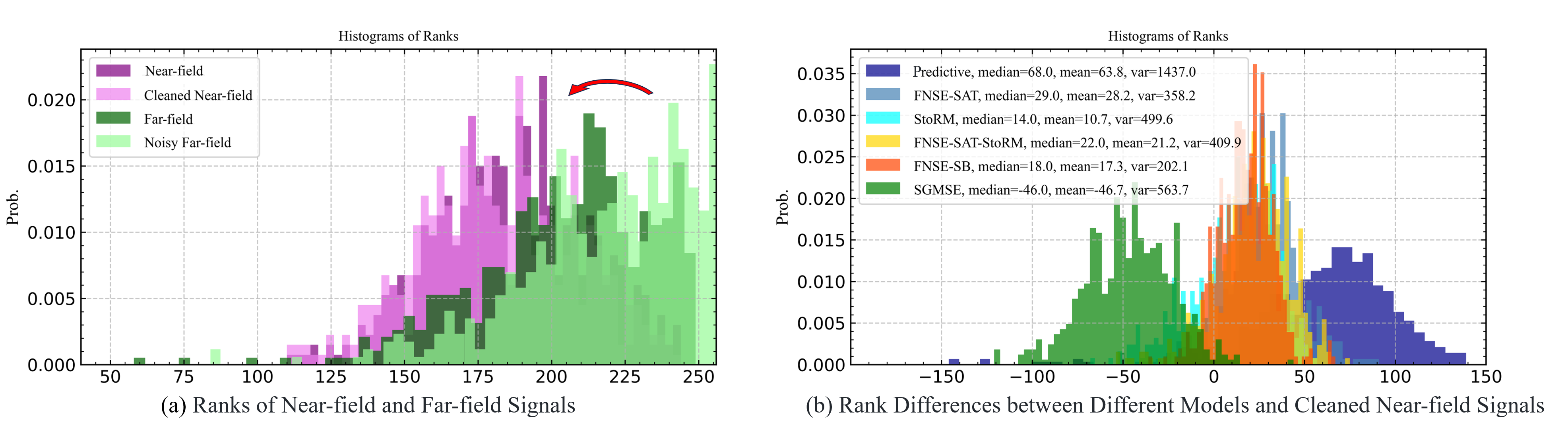}
        \caption{The ranks of both near-field and far-field signals are analyzed, along with the differences between the model outputs and the cleaned near-field signals. The term ``Cleaned near-field" refers to the enhanced signals obtained from the near-field, while ``Noisy far-field" denotes the far-field signals that have been intentionally corrupted with added noise. For the purpose of rank calculation, an absolute threshold $\eta$ of 0.5 has been set.}
        \label{fig:rank}
\end{figure*}

\subsubsection{Comparisons with SOTA methods}



Table~\ref{tab:baseline-result} presents objective comparisons on the real-far test set, revealing several key observations. First, although predictive methods without any generative capability can improve CER to some extent, they perform poorly on all other intrusive and non-intrusive metrics, even with data augmentation. Second, for the ``FNSE-SAT" method, which relies on supervised adversarial training and generative adversarial networks, data augmentation significantly enhances model performance. Third, for the two-stage ``SToRM" approach, which combines a predictive model with a diffusion model, the quality of the output from the first-stage predictive model, serving as a prior for the second-stage score-matching diffusion, largely determines the final generation quality. In the ``SToRM" framework, the predictive model utilizes the original NCSN++ architecture, which uniquely processes the spectrogram by treating the time and frequency dimensions isotropically, akin to how an image is handled in visual processing networks. This contrasts with typical audio processing models~\cite{wang2023tf,le2022inference}, where the time and frequency dimensions are often treated distinctly and modeled separately. In contrast, ``FNSE-SAT-SToRM" replaces the predictive model with a fixed-parameter and data-augmented ``FNSE-SAT". Compared to ``FNSE-SAT", although the objective metrics of the latter show improvement, the CER metric slightly decreases, indicating that the second-stage diffusion model generates semantic modifications. 

Furthermore, among all evaluated single-stage diffusion models, the SB approach demonstrates superior performance on the far-field to near-field speech enhancement task across various probabilistic sampling methods. Notably, through the integration of GANs and model scaling techniques, the proposed framework achieves a remarkable reduction in CER by up to 14.58\% compared to the original far-field signal, while maintaining enhanced speech quality.

\begin{table*}[!t]
	\caption{MUSHRA scores among different methods on the Test set. The confidence level is 95\%, and we performed a t-test comparing FNSE-SBGAN with FNSE-SAT-SToRM. Note that $^{**}p<0.05$, $^{*}p<0.1$.}
        \vspace{-16pt}
	\centering
	\Huge
	\resizebox{1.0\textwidth}{!}{
		\begin{tabular}{c|cccccccc}
			\hline
			Models &Cleaned Near-field &Predictive&SGMSE&FNSE-SAT&SToRM &FNSE-SAT-SToRM&FNSE-SBGAN  &FNSE-SBGAN-Large \\
                \hline
			MUSHRA &93.43$\pm$0.94&79.82$\pm$1.21&80.33$\pm$1.35&87.29$\pm$1.11&81.97$\pm$1.32&87.65$\pm$1.07&**89.09$\pm$0.96&\textbf{89.97$\pm$0.93}\\
			\hline
	\end{tabular}}
	\label{tab:all-baseline-mushra}
\end{table*}

\begin{figure*}[!t]
        \centering
        \vspace{-12pt}
        \includegraphics[width=1.0\linewidth]{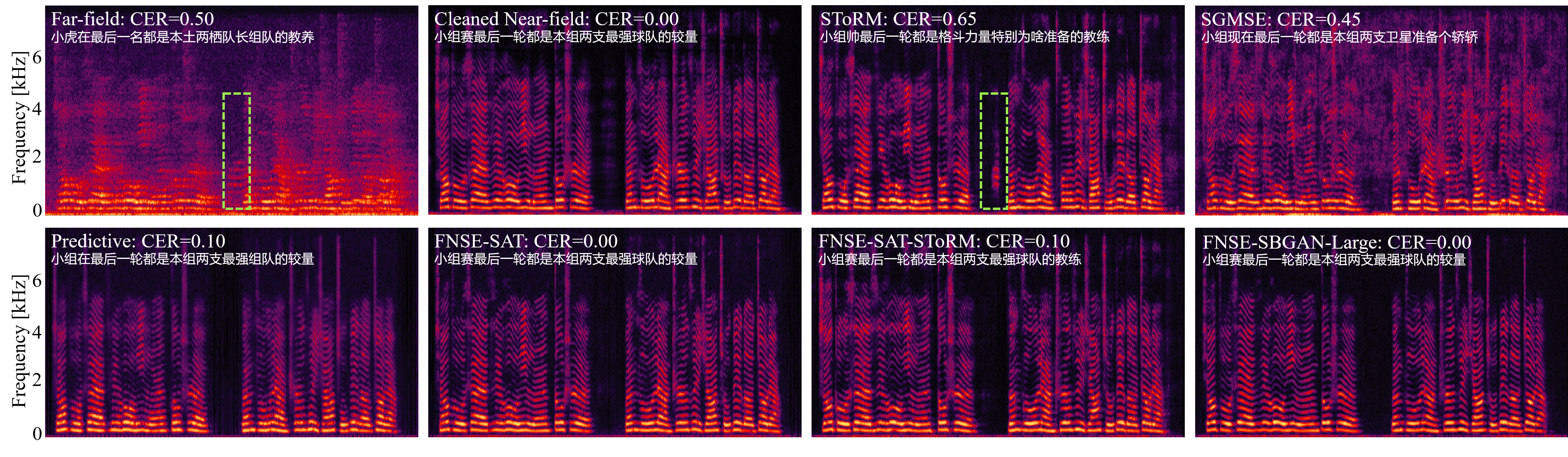}
        \vspace{-20pt}
        \caption{Spectral visualizations from an indoor sample. The text in the images includes the model names, the corresponding Chinese speech recognition results for the samples, and the associated CER.}
        \label{fig:demo-indoor}
\end{figure*}

\subsubsection{Rank Analysis}
\label{subsub:rank-analysis}
The phenomenon of rank expansion as described in Eq.~(\ref{eq:denoise-rank}) is visually demonstrated in Fig. \ref{fig:rank}(a). Here, signals normalized in energy in the time domain have undergone a 512-point STFT analysis, resulting in 257 frequency bins. Notably, a red arrow indicates a reduction in rank corresponding to the suppression of noise and reverberation. In this context, matrix rank is primarily associated with singular values, which more accurately characterizes the energy distribution in non-square matrices.

The rank differences are calculated by subtracting the ranks of the far-field test set processed with various baseline methods from those obtained by the ``Cleaned near-field" samples. These differences are statistically represented in Fig. \ref{fig:rank}(b). The distribution of medians from left to right suggests a decreasing trend in the rank values of the model outputs. A median and mean closer to zero, along with a smaller variance, indicate a closer match between the model's output rank and the target ``Cleaned near-field" rank.

Integrating the statistical results from Fig. \ref{fig:rank}(b) with the performance metrics in Table~\ref{tab:baseline-result}, it is evident that methods with medians and variances closer to zero, and particularly those with smaller variances, tend to perform better on this task. Most methods result in rank differences greater than zero for the majority of processed samples, except for ``SGMSE," which tends to generate components that increase the rank. This analysis helps in identifying the effectiveness of different methods in approximating the rank structure of the ``Cleaned near-field" target, thereby providing insights into their noise and reverberation suppression capabilities.

\begin{figure*}[!t]
        \centering
        \includegraphics[width=1.0\linewidth]{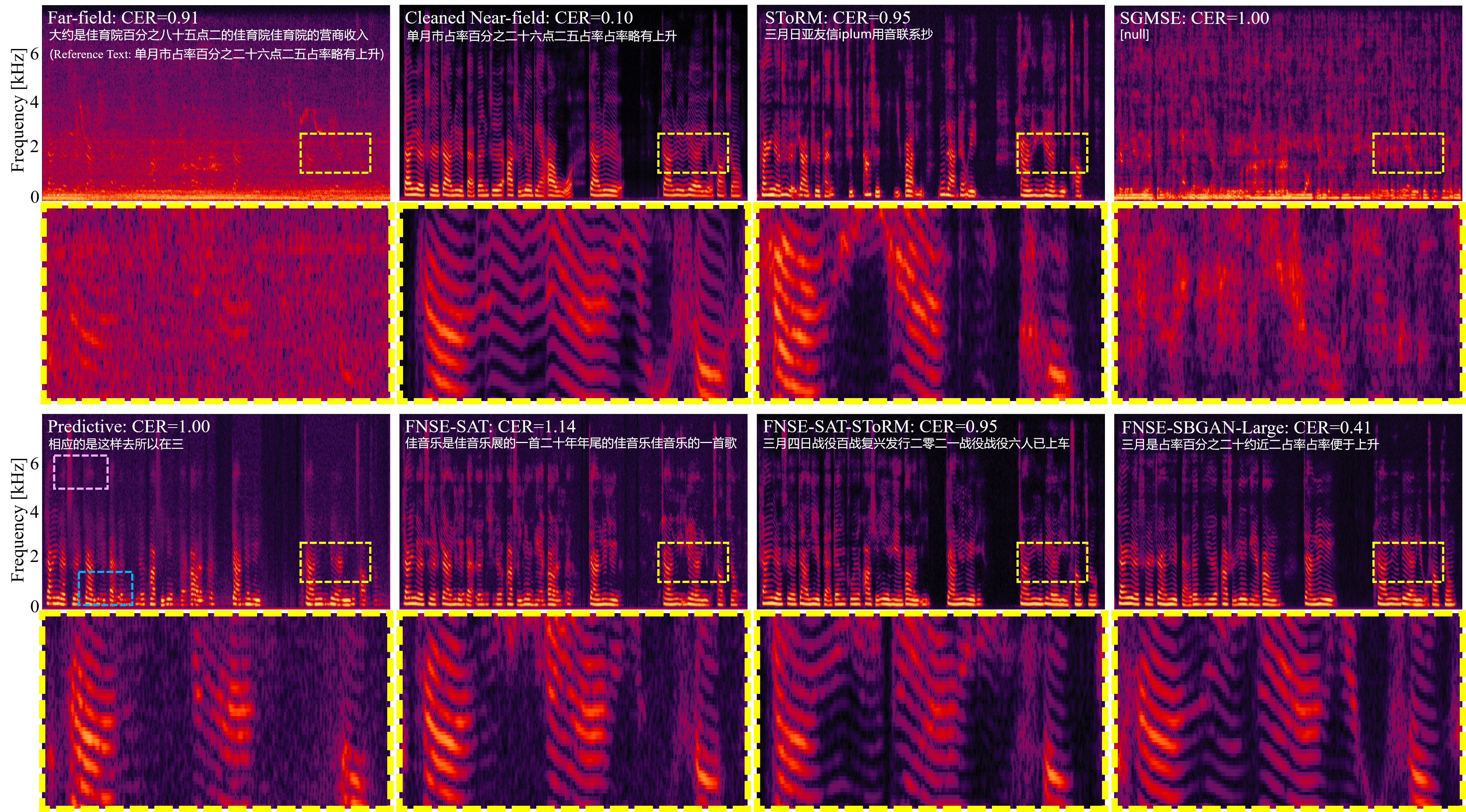}
        \vspace{-16pt}
        \caption{Spectral visualizations from an outdoor sample. The images in the second and the fourth row represent zoomed-in views of the corresponding positions in the previous row.}
        \label{fig:demo-outdoor}
\end{figure*}

\begin{figure}[!t]
        \centering
        \includegraphics[width=1.0\linewidth]{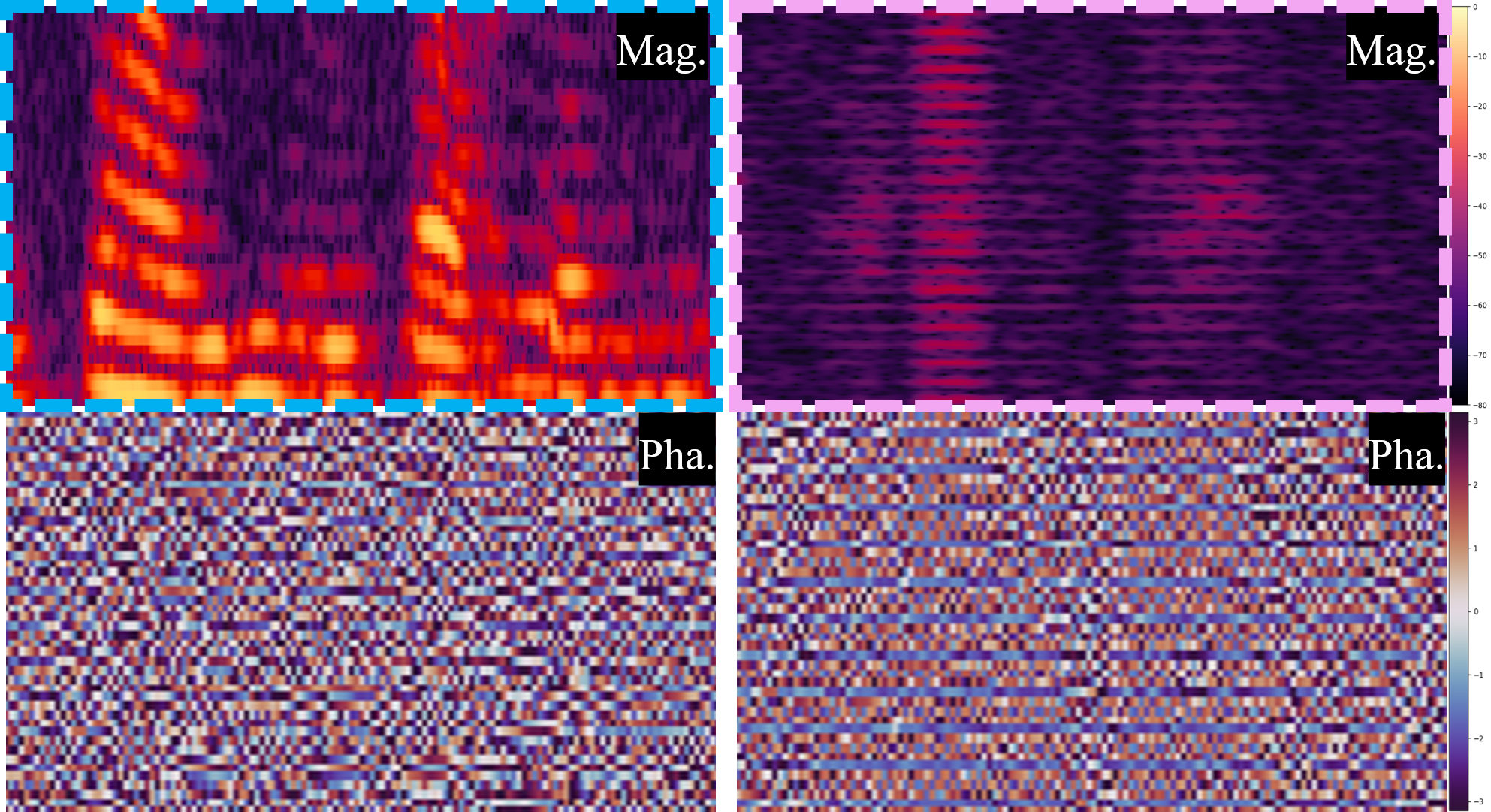}
        \vspace{-16pt}
        \caption{The zoomed-in views of the magnitude and phase spectrums of the Predictive model from Fig.~\ref{fig:demo-outdoor}.}
        \label{fig:demo-outdoor-phase}
\end{figure}

\subsubsection{Subjective Evaluations}
For subjective evaluations, we employ the MUSHRA (MUltiple Stimuli with Hidden Reference and Anchor) testing methodology as outlined in ITU-R BS.1534-3~\cite{series2014method}, utilizing the BeaqleJS platform~{\cite{kraft2014beaqlejs}}. A total of 17 native Chinese speakers, all specializing in audio signal processing, are involved in the testing. In the MUSHRA test, participants evaluate speech processed by various algorithms on a scale from 0 to 100, assessing overall similarity to the cleaned near-field reference, with a particular emphasis on semantic similarity.

Table~\ref{tab:all-baseline-mushra} presents the MUSHRA results, which reveal that our FNSE-SBGAN is statistically superior to FNSE-SAT-SToRM ($p < 0.05$) and other baselines, further demonstrating the advantage of our method in achieving subjective quality close to the ground truth signal. The FNSE-SBGAN-Large model with a larger number of parameters exhibits better subjective performance.

\subsubsection{Spectral Visualizations}
Fig.~\ref{fig:demo-indoor} presents spectral visualizations of an indoor sample with a high level of reverberation. The far-field recognition results, which show a CER of 0.50, clearly indicate that high reverberation significantly degrades ASR model performance. While the directly trained ``Predictive" model demonstrates decent noise reduction capabilities on this sample, it falls short in recovering mid-high frequency details and restoring energy. In contrast, the pure diffusion method ``SGMSE" and the two-stage joint method ``SToRM" (which combines predictive and diffusion approaches) achieve better compensation for mid-high frequency energy loss, even though they do not employ audio-specific architectures like TF-GridNet. The inference results of ``SGMSE" reveal artificial generation artifacts that induce semantic modifications. Notably, the ``SToRM" method's first-stage NCSN++ network, which treats time-frequency dimensions isotropically, struggles to suppress unexpected harmonic structure noise in the background scene, as highlighted in the green boxes. This limitation is effectively addressed in the ``FNSE-SAT-SToRM" variant through its first-stage FNSE-SAT module based on TF-GridNet, which demonstrates superior interference suppression. Our proposed ``FNSE-SBGAN-Large," while also built on the NCSN++ diffusion framework, achieves enhanced mid-high frequency spectral energy compensation and avoids semantic modification. 

Fig.~\ref{fig:demo-outdoor} presents spectral visualizations of an outdoor sample captured in ultra-far-field conditions (greater than 12 meters) with strong noise interference, resulting in a far-field CER of 0.91 that significantly alters the semantic content. For this low-SNR sample, both the ``Predictive" and ``SToRM" models exhibit over-suppression artifacts. The ``SGMSE" method completely fails to recover any useful information. While ``FNSE-SAT-SToRM" and ``FNSE-SAT" appear to preserve and recover the target spectra, their artificial generation artifacts substantially distort the original semantics. Our proposed ``FNSE-SBGAN-Large" generates harmonic information while maximally preserving the semantic content. Detailed observations from the zoomed-in views (yellow boxes) further confirm its recovery advantages. Note that the speaker deviates slightly from the reference text, resulting in a CER that is not equal to zero, even in near-field conditions.

Additionally, we offer insights into the frequency-domain striping phenomenon discussed in~\cite{LEI:FNSESAT}, particularly observed in the directly tuned ``Predictive" model. Fig.~\ref{fig:demo-outdoor-phase} presents zoomed-in views of the magnitude and phase spectra (blue/pink boxes from Fig.~\ref{fig:demo-outdoor}), revealing a \textit{regression to the mean} effect~\cite{whang2022deblurring} that manifests as frequency-domain striping, especially pronounced in high frequencies. From a rank analysis perspective, this issue arises from the ``Predictive" model's limited generative capacity: for mid-high frequency components corrupted in far-field signals, it produces only coarse frequency-dependent energy estimates. This results in over-suppression and striping in the magnitude spectra, along with non-physical phase fluctuations and insufficient temporal variation in the phase spectra. Consequently, its processed outputs exhibit substantially lower ranks compared to clean near-field targets, leading to the largest mean/median rank differences, as shown in Fig.~\ref{fig:rank}(b). In contrast, the ``SGMSE" method, which tends to reconstruct signals from white noise, generates excessive artificial components, typically resulting in higher ranks than those of clean targets.

The audiometry samples are available at \url{https://github.com/Taltt/FNSE-SBGAN}.

\section{\label{sec:conclusions}Conclusions}
In this study, we addressed the significant mismatches between simulated and real-world acoustic data by revisiting the single-channel far-field to near-field speech enhancement (FNSE) task. We proposed FNSE-SBGAN, a framework that integrates a Schrödinger Bridge (SB)-based diffusion model with generative adversarial networks (GANs). Our approach achieved state-of-the-art performance across various metrics and subjective evaluations, significantly reducing the character error rate (CER) by up to 14.58\% compared to far-field signals. Experimental results demonstrated that FNSE-SBGAN preserves superior subjective quality. To systematically analyze model behavior, we introduce a time-frequency matrix rank analysis framework, revealing distinct spectral recovery characteristics of generative and predictive approaches. This analysis further reinterprets frequency-domain striping artifacts in predictive models as rank deficiencies caused by spectral averaging effects. The proposed methodology establishes a new benchmark for real-world far-field enhancement while providing interpretable insights into model capabilities and limitations.

\end{document}